\documentclass[aip,jcp,reprint]{revtex4-1}

\usepackage{subfigure}
\usepackage{amsmath}
\usepackage{graphicx}
\usepackage{color}

\usepackage[usenames,dvipsnames]{xcolor}
\usepackage[normalem]{ulem}

\begin{document}
\title{Molecular dynamics simulations of shear-induced thermophoresis and non-Newtonian flow in compressible fluids}
\date\today

\newcommand\dop{\affiliation{Department of Physics and Institute of 
Nanotechnology and Advanced Materials,
Bar-Ilan University, Ramat Gan 52900, Israel}}

\newcommand\nyu{\affiliation{NYU-ECNU 
Institutes of Physics and Mathematical Sciences at 
NYU Shanghai, 3663 Zhongshan Road North, Shanghai, 200062, China}}

\author{Madhu Priya$^*$}\email{madhuatjnu@gmail.com}\dop
\author{Yitzhak Rabin}\email{yitzhak.rabin@biu.ac.il}\dop\nyu

\begin{abstract}
We use molecular dynamics simulations to study the behavior of a compressible Lennard-Jones fluid in simple shear flow
 in a two-dimensional nanochannel. The system is equilibrated
in the fluid phase close to the triple point at which gas,
liquid and solid phases coexist and is subjected
to steady shear in Couette geometry. It is observed that at higher shear 
rates, the system develops a density gradient perpendicular to the direction 
of flow and exhibits solid-like layering near the boundaries. Both the number of solid-like layers 
and the number of layers that move with the velocity of the neighboring wall, increase with the shear rate. 
We argue that the inhomogeneous density profile develops as the consequence of thermophoresis due to the non-uniform temperature profile
produced by shear-induced viscous heating in the simulated flow cell. The above phenomena
are accompanied by non-Newtonian effects such as nonlinear velocity profiles, 
inhomogeneous stress distributions and  shear rate dependent viscosity
which exhibits shear thinning followed by shear thickening as the shear rate is increased. 
The connection between these phenomena is discussed.
\end{abstract}

\maketitle

\newpage

\section{Introduction}
\label{Introduction}
Molecular dynamics (MD) is an important tool for 
investigating properties of a fluid under flow and is often used to explore
systems under conditions which are difficult to achieve and control in experiments, e.g., 
flows at high shear rates in nanochannels 
\cite{Jabbarzadeh2000}.
The boundary conditions (BC's) for such computer 
experiments are very important, 
especially for systems of nanoscale dimensions in which properties of the wall-fluid interface
have a significant effect on the flow. Some of the earlier 
MD simulations of Couette flow considered smooth walls and reported 
wall slip \cite{Trozzi1984}. However, a more recent study showed that for wetting liquids, slip 
arises only at very high shear rates at which the response of the fluid to the applied 
shear is no longer linear \cite{Barrat1994}.
Some authors introduced the no-slip BC's explicitly,
i.e., they assumed that the fluid layer next to the wall moves with the 
velocity of the wall, but subsequent studies have shown that 
imposition of such BC's is incorrect since whether the fluid particles will slip at 
the wall or not, depends on the strength of the wall-fluid interaction 
\cite{Thompson1990, Thompson1997}. 

Along with implementing the correct BC's, it is 
important to control the temperature in the system by choosing 
a thermostat which closely mimics experiment and is computationally efficient. 
Recent computer simulation studies showed 
that the choice of a thermostat has major effects on fluid flow
at high shear rates in confined channels  \cite{Bernardi2010,Yong2013a}. 
The authors considered several scenarios: (1) thermostating the wall (TW), 
(2) the fluid (TF) and (3)  both the wall and the fluid 
(TWTF). In TW the walls are made up of  
particles which are tethered by springs to their equilibrium lattice positions. 
TW simulations can reproduce the temperature profiles in actual experiments 
where the extra heat due to shear is dissipated through the walls. However,
since the wall particles are oscillating around their mean positions on some characteristic time scale, 
the effective roughness of the walls depends on the applied shear rate and therefore
TW simulations fail to describe systems in which  
wall roughness does not depend on the flow rate. In order to maintain constant wall
roughness a possible choice is to use TF. This can be done 
either by assuming a linear velocity profile (profile-biased thermostat), 
or by measuring the actual velocity profile obtained in the simulation
and implementing the thermostat using this profile (profile-unbiased thermostat). 
A profile-biased thermostat is limited in its accuracy since in many cases the linear velocity profile 
assumption breaks down at high shear rates. 
While this problem can be solved by using a profile-unbiased thermostat,  TF thermostats cannot reproduce the 
experimental conditions in which only the walls are thermostated. Finally, TWTF 
simulations are computationally expensive and tend to distort the
effects of viscous heating on fluid dynamics that become increasingly important at higher shear rates.

In the present work, we use MD simulations to study the effect 
of simple shear on
a fluid of monodisperse particles that interact with each other via a 
Lennard-Jones (LJ) potential and are confined in 
a two-dimensional (2D) nanochannel. In order to amplify the effects of shear on
the temperature and density profiles in the nanochannel, we study this system in the region 
of the phase diagram where the compressibility is large, i.e., at the triple 
point density and
at temperature that is slightly higher than that of the  liquid-solid transition.  Steady shear
is applied by moving the upper wall with a constant velocity while the lower wall remains at 
rest throughout the simulation.  
The walls are made of particles which are fixed with 
respect to each other and therefore the roughness of wall remains constant during the simulation. 
We thermostat two layers of fluid particles next to each wall  
\cite{Rapaport1995, RapaportDis}, and therefore allow 
temperature gradients to develop between the walls and the bulk of the fluid
at high shear rates where shear-induced heating becomes important. 

The paper is organized as follows. We  provide details of our simulation 
setup in Sec. \ref{theory}. In Sec. \ref{results} we present the calculated density, flow and temperature 
profiles, discuss the connection between these results and thermophoresis in temperature gradients
and analyze the various non-Newtonian characteristics of flow at high shear rates. We conclude 
the paper by summarizing and discussing our results in Sec. \ref{discussion}.

\section{Simulation details}\label{theory}
\noindent  The fluid particles interact 
with each other via the
LJ potential,
\begin{equation}
U_{LJ}(r)=4\epsilon\Big[\Big(\frac{\sigma}{r}\Big)^{12}-\Big(\frac{\sigma}{
r }
\Big)^ { 6 } \Big ],
\end{equation}
which is terminated and shifted at $r=r_{\rm cut}=2.5\sigma$, so that the 
truncated potential $\bar{U}_{LJ}(r)$ is defined as,
 \[ \bar{U}_{LJ}(r) =
  \begin{cases}
    U_{LJ}(r) - U_{LJ}(r_{\rm cut}) & \quad {\text {if}}~ r < r_{\rm cut} \\
        0 & \quad {\text {if}}~ r \ge r_{\rm cut}\\
  \end{cases}
\].

We use reduced LJ units in which the interaction parameter $\epsilon$, 
the mass $m$ and length scale $\sigma$ are taken
to be unity (the Boltzmann constant is taken as unity as well). The simulations 
are performed in $NVT$ 
ensemble. The dynamics is solved by using a velocity-Verlet integrator with a 
time step of 
$\delta t = 0.005 \tau_{LJ}$, where $\tau_{LJ} = \sigma(m/\epsilon)^{1/2} = 1$ 
is the LJ time unit. Most of the results were obtained at triple point density 
(0.694) that corresponds to $N = 1600$ 
particles  in an area of $48\sigma\times 48\sigma$
 (other densities were obtained by 
changing the number of particles at fixed volume of the 
system). The nanochannel is 
constructed by placing two parallel solid walls at $y=0$ and $y=L_{y}$, where 
$L_{y}=48\sigma$ is the distance between the walls. Each of the walls is made 
of $43$ particles 
with centers located at positions $y=0$ and $y=48\sigma$, respectively, such 
that the
horizontal separation between the wall particles is $1.12\sigma$, which corresponds 
to the minimum of the LJ potential. The wall-fluid particle interactions are the same as 
between fluid particles. Periodic boundary conditions with period 
$L_x = 48\sigma$ along the $x$ direction are imposed. 

We start the simulation from a configuration 
in which the fluid particles are placed on a square lattice,  between the two solid 
walls. The initial velocity of the $i$th fluid particle is chosen from a 
Maxwell-Boltzmann distribution at temperature $T$, to which we add a 
velocity given by the product of the $y$-position of the particle and the shear 
rate where the latter is defined by the constant velocity of the 
upper wall $U\hat{x}$ as
$\dot{\gamma}=U/ L$. 
The distance between the walls is kept fixed during the simulation.

The fluid particles obey the following equations of motion,
\begin{eqnarray}
 &&{\textbf v}_{i}\Big(t+\frac{\Delta t}{2}\Big) = {\textbf v}_{i}(t)+ 
\frac{\Delta 
t}{2}{\textbf F}_{i}(t),\\
&&{\textbf r}_{i}(t+\Delta t) = {\textbf r}_{i}(t) +  
{\textbf v}_{i}\Big(t+\frac{\Delta t}{2}\Big)\Delta t,\\
&&{\textbf v}_{i}\Big(t+\Delta t\Big) = {\textbf v}_{i}\Big(t+\frac{\Delta 
t}{2}\Big)+ \frac{\Delta t}{2}\textbf F_{i}\Big(t+\frac{\Delta t}{2}\Big).
\end{eqnarray}
In the above set of equations ${\textbf v}_{i}(t)$ and ${\textbf F}_{i}(t)$ represent 
the instantaneous velocity of particle $i$ and the instantaneous force acting on it, respectively.

In order to define a local instantaneous temperature $T(y,t)$, we divide the 
system into $40$ bins along $y$-axis such that each bin contains approximately a single layer
of particles. The average velocity of particles in bin $\alpha$ is defined as 
\begin{equation}
\label{streamVel}{\textbf u}_{\alpha}(t) = 
\frac{1}{N_{\alpha}}\sum_{i=1}^{N_{\alpha}}{\textbf v}_{i}(t),
\end{equation}
where we sum over the instantaneous velocities of the $N_{\alpha}$ particles in this bin. We define the 
peculiar velocity of particle $i$ in this bin as  \cite{Loose1992}
\begin{equation}
 {\textbf v_{i}^{p}}(t) = {\textbf 
v}_{i}(t)-{\textbf u}_{b}(t).
\end{equation}
Note that the peculiar velocity is defined in the rest frame of the average particle in the bin and 
can therefore be used to define the local temperature in the $\alpha$th bin as
\begin{equation}
T_{\alpha}(t) = \frac{\sum_{i}^{N_{\alpha}}
 [{\textbf 
v}_{i}^{p}(t)]^2}{2(N_{\alpha}-1)}.
\end{equation}
where, in the denominator, we subtracted $2$ from the number of degrees of freedom ($2N_{\alpha}$) in the $\alpha$-th bin 
because they were already included in the definition of local streaming 
velocity, Eq. (\ref{streamVel}). 
A constant temperature ($T$) is maintained near the 
walls by thermostating two liquid layers adjacent to each wall ($\alpha=1,2$ and $\alpha=39,40$, respectively) using velocity rescaling. 
At each time step 
the velocities of the particles in the above four bins are rescaled by a factor $\lambda$ 
defined in terms of the 
instantaneous temperature $T_{\alpha}(t)$ and the 
fixed temperature $T$ as  
$\lambda=\sqrt{(T/T_{\alpha}(t))}$. The above procedure allows us to maintain simultaneously
constant roughness of the walls and a physically
realistic temperature distribution inside the simulation box.

\section{Results}\label{results}

\begin{figure}[ht!]
     \begin{center}
        \subfigure[ ]{%
           \includegraphics[width=0.21\textwidth]{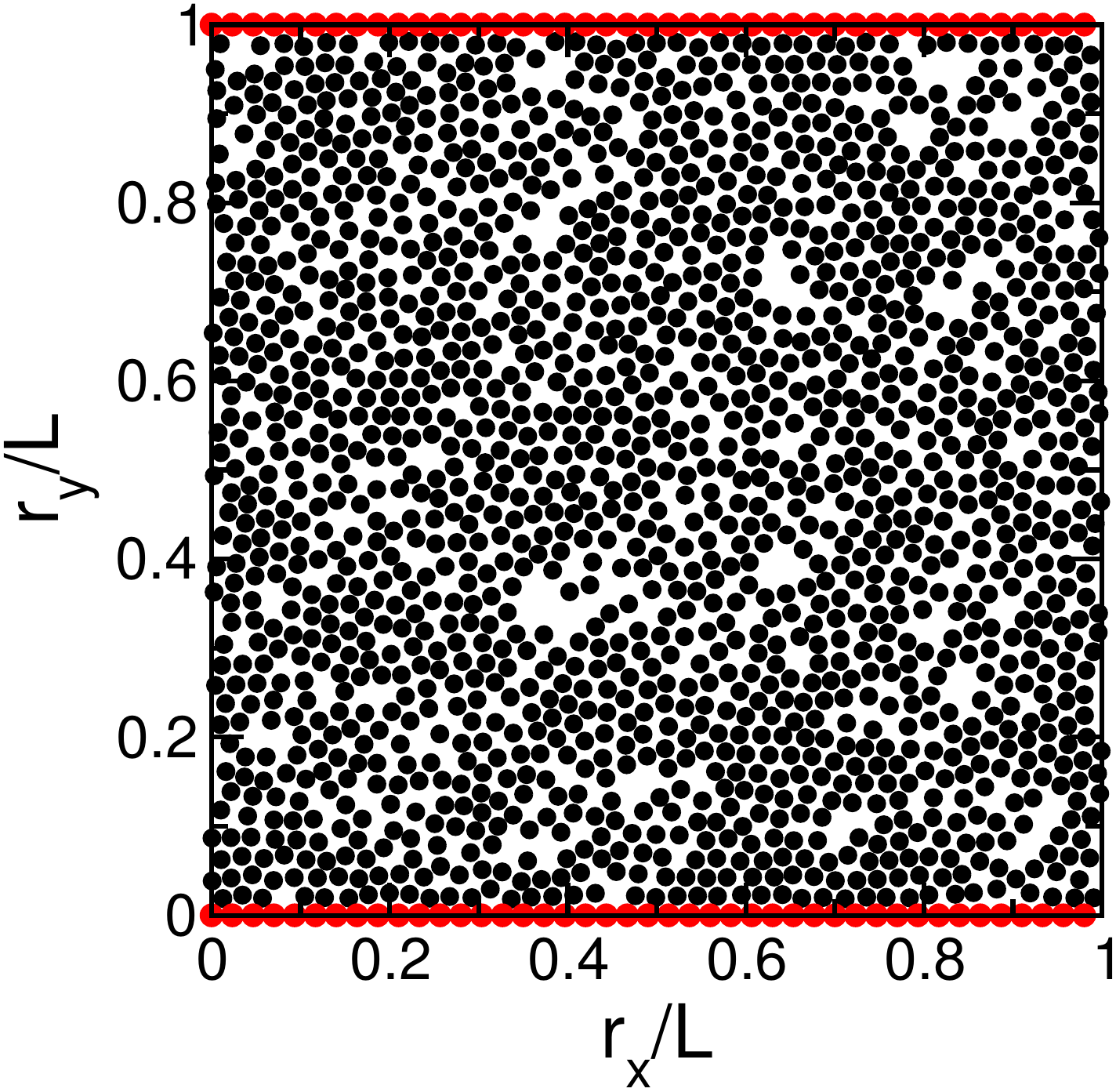}
           {\label{snapshotPZero}}
        } \hspace*{0.1cm}
        \subfigure[ ]{%
           \label{snapshotPThree}
           \includegraphics[width=0.21\textwidth]{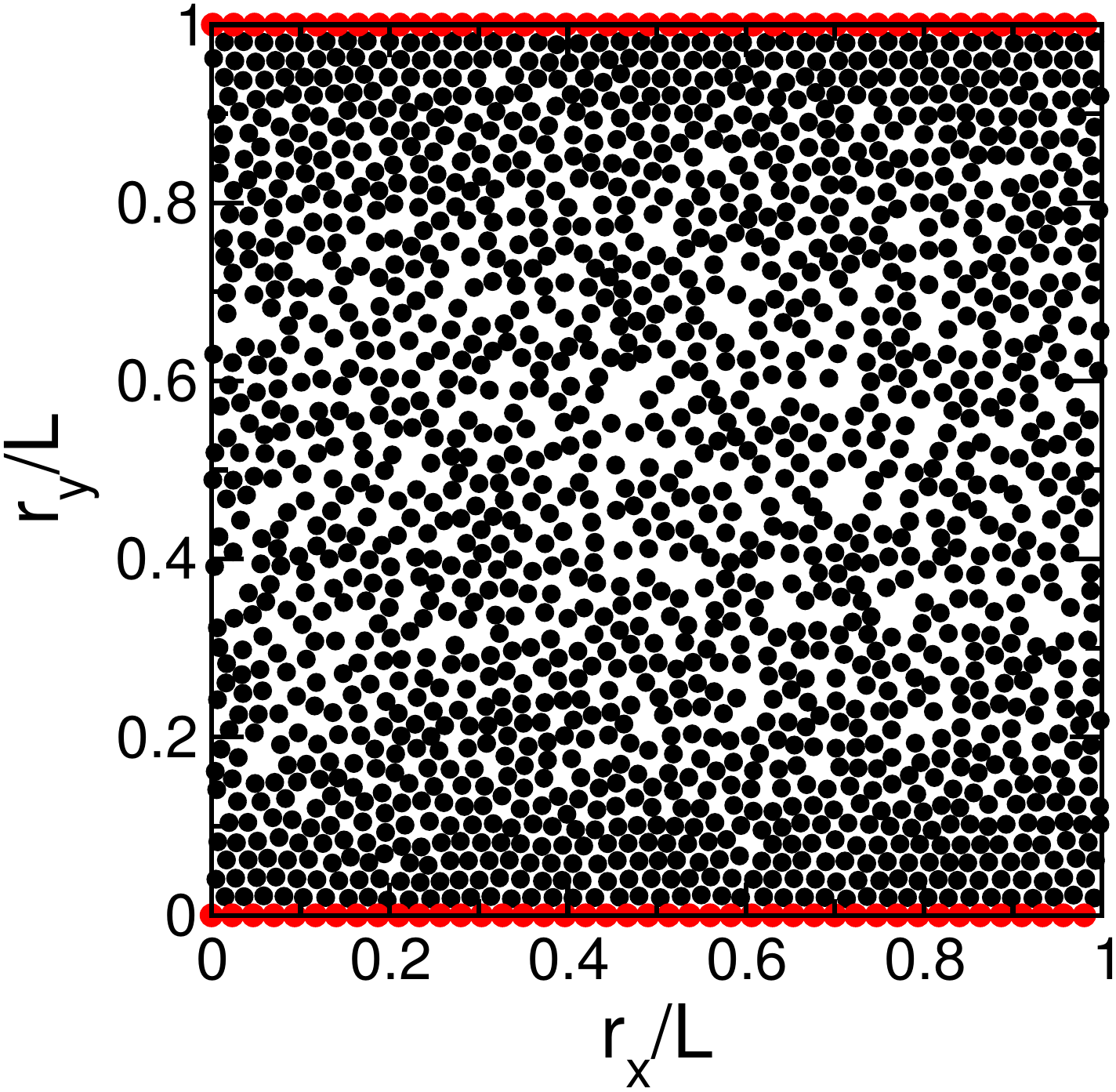}
        } 
    \end{center}
    \caption{%
        Snapshots of the system at $T = 0.44$ and $\rho=0.694$, where  
  wall particles and fluid particles are shown as red and black solid discs, respectively. 
(a) at equilibrium,
$\dot{\gamma}=0$ and (b) in steady shear, $\dot{\gamma}=0.3$. 
     }%
   \label{fig:subfigures}
\end{figure}

Typical snapshots of the system in equilibrium 
$\dot{\gamma}=0$ and under strong shear $\dot{\gamma}=0.3$,  are shown in 
Figs.~\ref{snapshotPZero} and ~\ref{snapshotPThree}, respectively. Even though strong
density fluctuations are observed in both figures, inspection of Fig.~\ref{snapshotPZero} shows 
that (with the exception of $1-2$ fluid layers near the walls where some ordering is visible) 
the average density is uniform 
across the system in equilibrium. This is not the case in the high shear limit where the steady state 
density is minimal at the center of the system ($y=L/2$) and strongly increases towards the walls, 
Fig.~\ref{snapshotPThree}.
 
\begin{figure}[ht!]
\includegraphics[width=.45\textwidth]{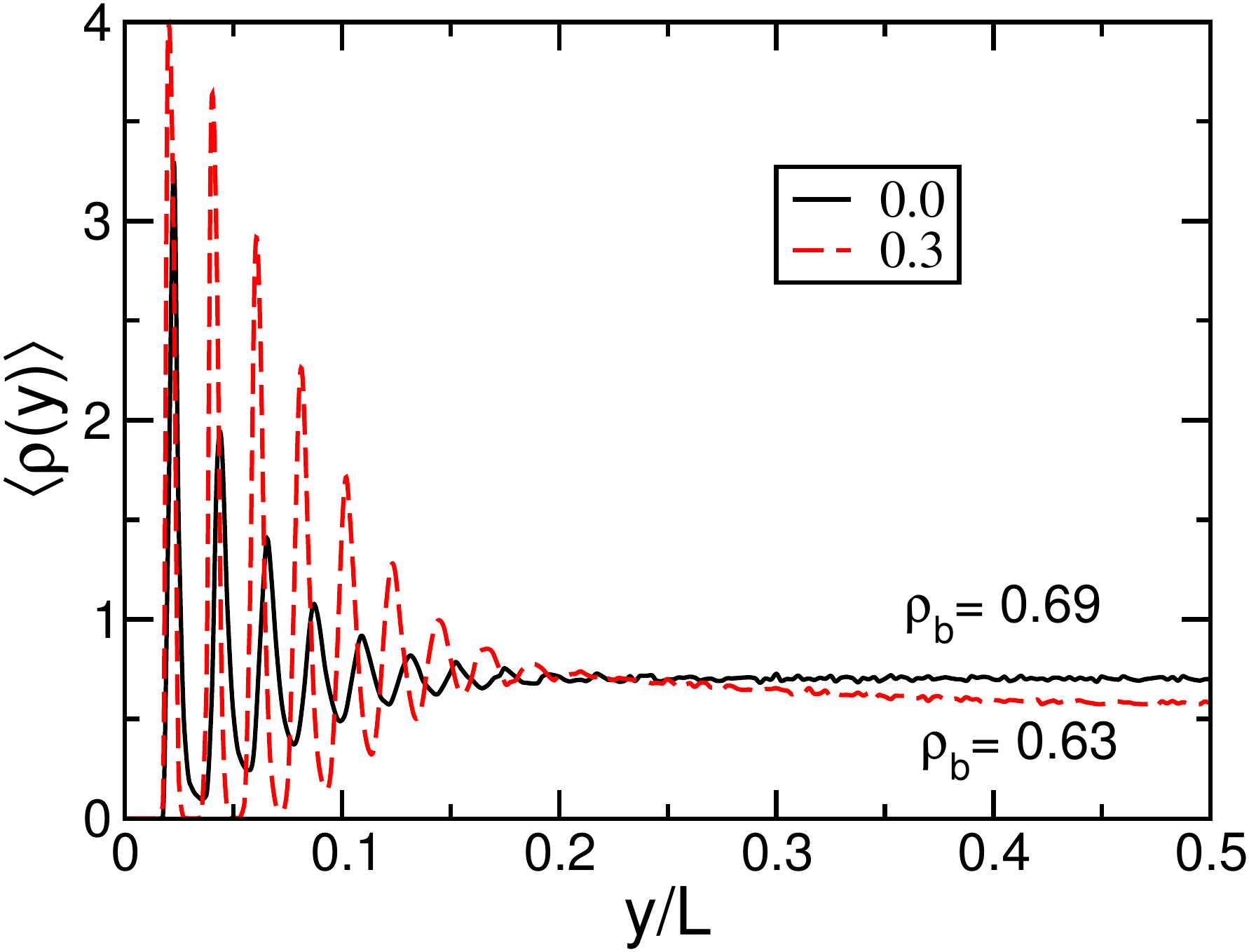}
\caption{\label{densityProfile}
 Comparison of $\langle\rho (y)\rangle$ at 
equilibrium and at steady state $\dot{\gamma} = 0.3$, for $T = 
0.44$ and $\rho = 0.694$.}
\end{figure}

In order to quantify the effect of steady shear on the density profile  
we divide the system into $400$ bins and average
the density in each bin over $x$ and over time. 
The resulting equilibrium and steady state profiles $\langle\rho (y)\rangle$ are plotted in Fig.~\ref{densityProfile}.
While density oscillations are clearly observed in both cases, the amplitudes of the peaks increase and their width 
and the separation between them decrease with shear rate, a signature of shear-induced solid-like layering near the
walls. The ratio of the amplitudes of the corresponding high shear rate and 
equilibrium peaks {\it increases} with distance from the 
walls, in agreement with the observation of shear-induced broadening of the solid-like boundary layers in
Figs.~\ref{snapshotPZero} and ~\ref{snapshotPThree}. Note that the enhancement 
and 
broadening of solid-like layering is accompanied by reduction of the bulk density $\rho_b$ in the center of the channel, to a
lower value ($0.63$) than the average density of the system ($0.694$).

\begin{figure}[ht!]
\includegraphics[width=.45\textwidth]{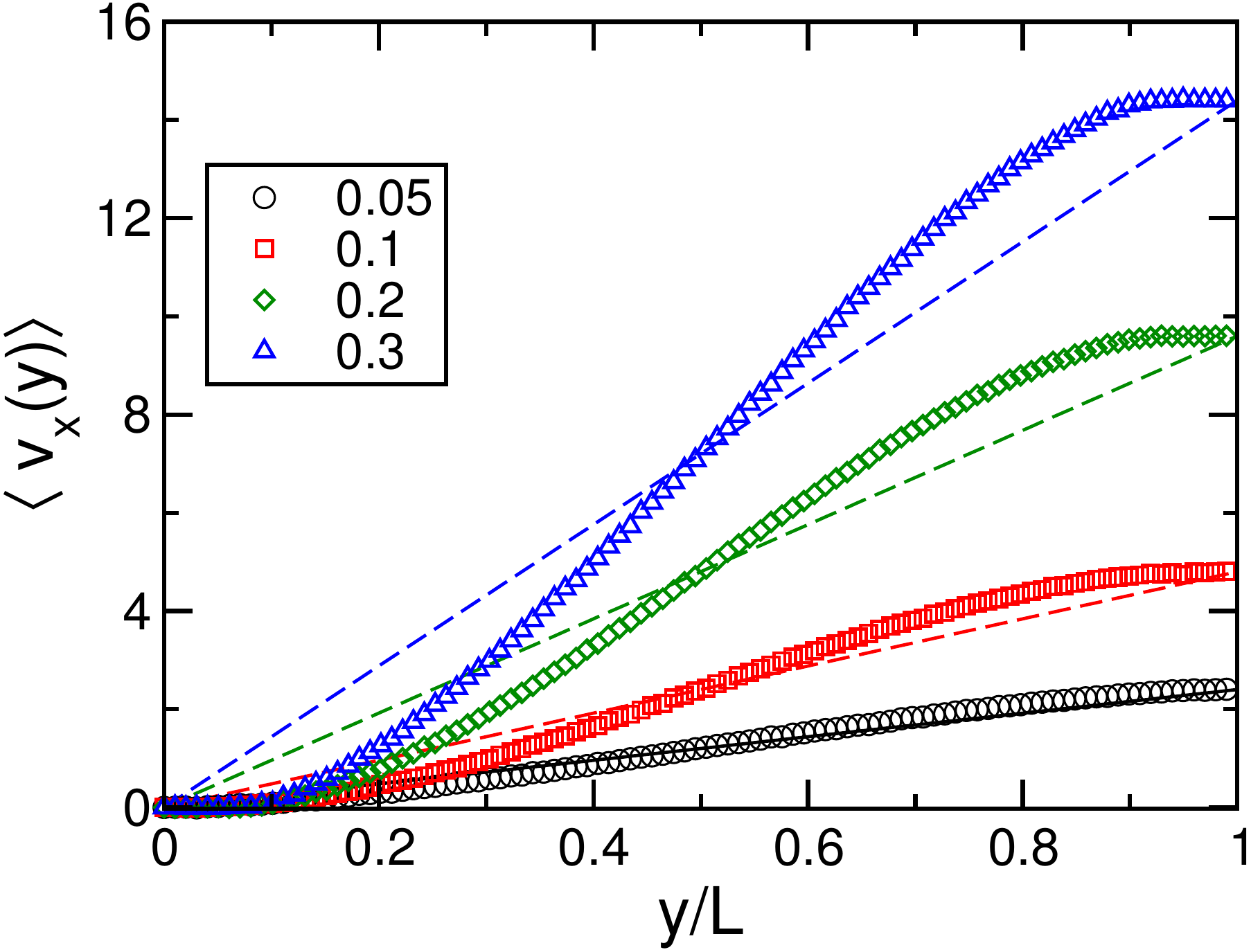}
\caption{\label{velocityProfile}
  Open symbols show average velocity of the particles along $x$-axis in 
different bins constructed according to their $y$-position for  
$\dot{\gamma}\ge0.05$. The dashed lines show a linear velocity profile 
 corresponding to each shear rate.}

\end{figure}

Having established the effect of shear on the density profile  we turn to examine its 
effect on the flow by measuring the y-dependence of the average velocity $\langle 
v_x(y)\rangle$  
for a range of applied shear rates (the average 
velocity of the particles in the $y$-direction vanishes, as expected on symmetry grounds). 
To this end we divide the system into $100$ bins along the $y$ direction
(we choose a lower number of bins as compared to the density 
measurements, to avoid bins with zero particles).   As shown in  
Fig.~\ref{velocityProfile}, around shear rate of $0.05$ one begins to observe
deviations from a linear velocity profile, $ v_x(y)=\dot{\gamma}y$.
These deviations 
manifest themselves in the formation of a boundary layer that moves 
together with the neighboring wall (and another boundary layer that remains at rest with respect to the stationary wall). This phenomenon 
has been previously observed in the case of strong wall-fluid interactions in a three 
dimensional LJ system and has been referred to as {\it locking} 
\cite{Thompson1990}. When the shear rate is further increased, the number of 
layers moving with the wall velocity increases and  
the velocity 
gradients in the bulk of the system increase as well beyond their nominal value ($\dot{\gamma}$).

\begin{figure}[ht!]
\includegraphics[width=.45\textwidth]{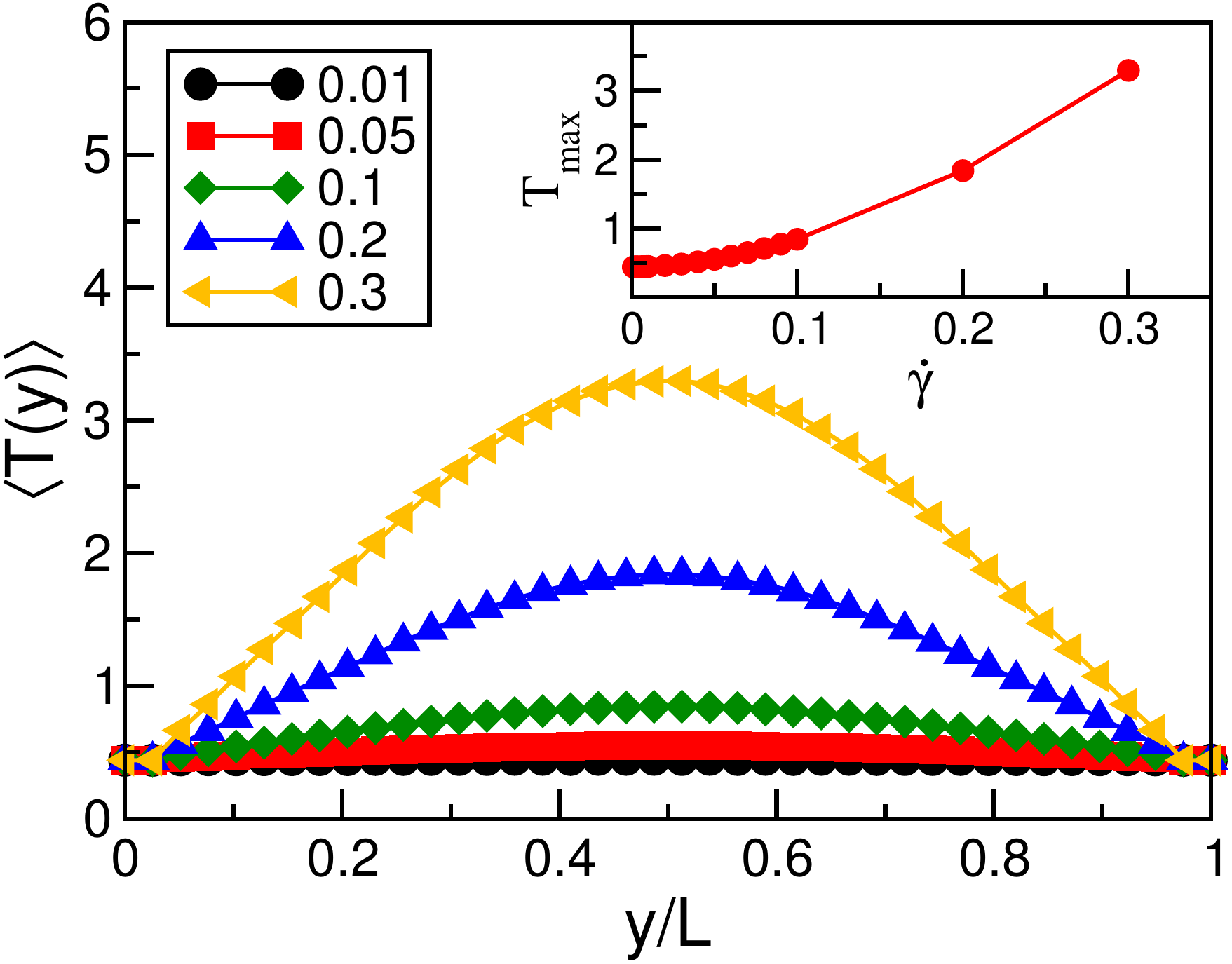}
\caption{\label{tempProfile}
Filled symbols show the temperature profile for various shear rates as 
mentioned in the legend. The lines are a guide to the eye. The inset shows the 
dependence of maximum of the temperature profile on the applied  
shear rate.   
}
\end{figure}

Since we would like to gain insight about the origin of the observed layering and locking phenomena
we proceed to examine the temperature profiles that develop in the system
with increasing shear rate. We find that the temperature profile is parabolic (in $y$),
with a maximum at the center of the system (the height of this maximum increases with 
shear rate as $\dot{\gamma}^2$ - see inset in Fig.~\ref{tempProfile}), and 
decreases to the nominal temperature $T$ 
at the two thermostated layers near each wall, as shown in Fig.~\ref{tempProfile}. Similar temperature profiles were also observed in 
other computer simulations of shear flow  \cite{Yong2013a,
Yong2013b}. This concurs with the expectation that shear-induced viscous heating
leads to higher temperature gradients, since the only way to
remove excess heat from the system is to increase these gradients in order to
enhance the diffusion of heat towards the thermostated walls.

\begin{figure}[ht!]
\includegraphics[width=.45\textwidth]{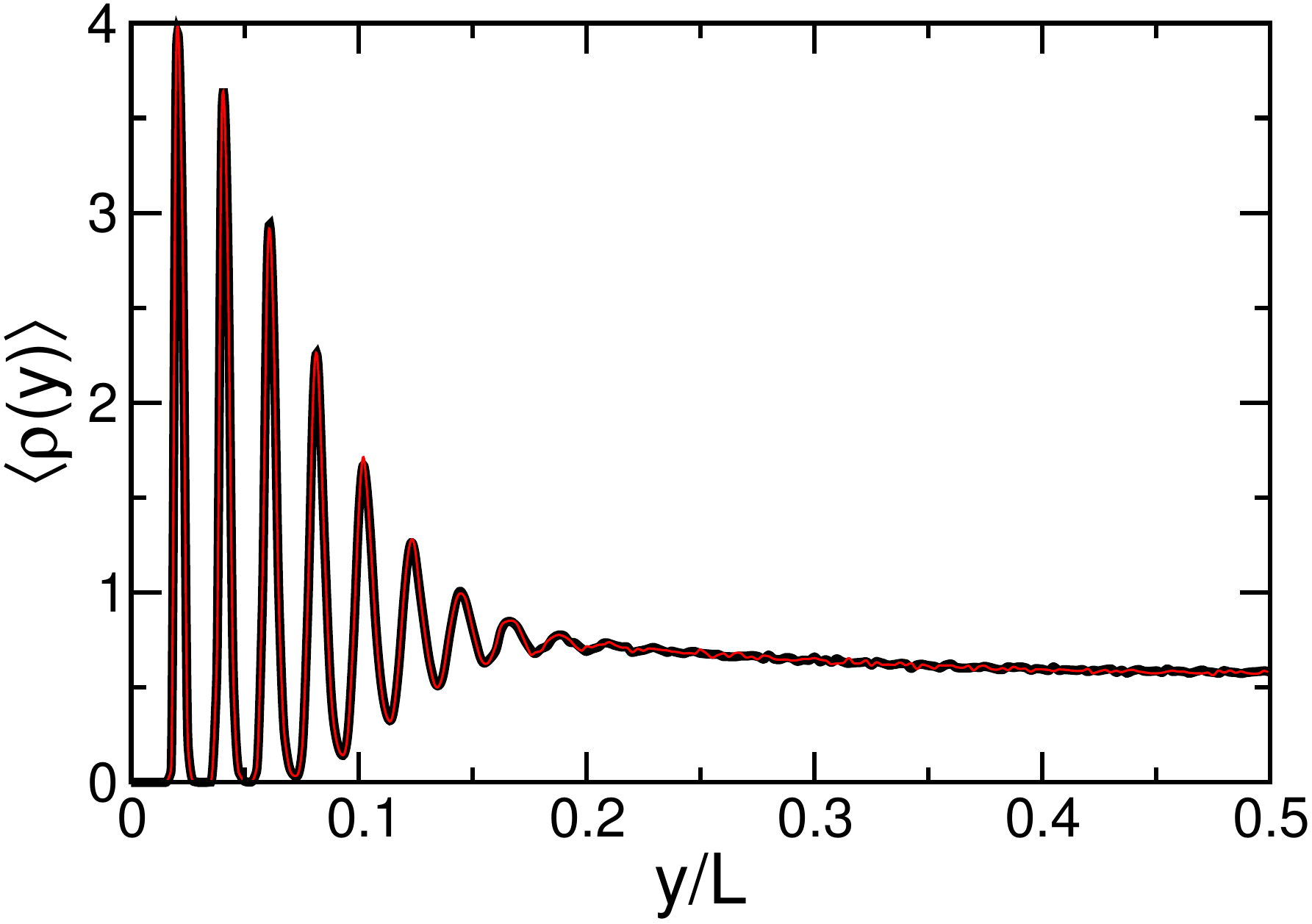}
\caption{\label{densityProfileComp}
 The density profile obtained for equilibrium fluid (black thick  
line) is compared to that obtained for $\dot{\gamma} = 0.3$
(red solid line), both with identical temperature profile.}  
\end{figure}

In order to check whether the temperature profiles completely determine 
the corresponding density profiles (at the same shear rates), in Fig.~\ref{densityProfileComp} we compare the 
steady state density profile of a sheared fluid with $\dot\gamma = 0.3$ to that 
of a fluid at rest but with an identical temperature profile. Since the resulting 
density profiles are indistinguishable, we conclude that shear-induced layering 
arises as the result of the coupling between density and temperature gradients in
the sheared fluid and thus the density depends on the shear rate through its effect on the temperature
profile, i.e., $\langle\rho(y)\rangle$ is a function of $\langle T(y)\rangle$ 
only.

In the absence of shear, the coupling between local temperature and  
concentration profiles gives rise to thermophoresis, also known
as the Ludwig-Soret effect \cite{Ludwig1856,Soret1880,Groot1984}. Although the 
Ludwig-Soret effect 
has been mostly 
studied in colloidal dispersions and binary mixtures 
\cite{Duhr2006a,Duhr2006b,Wuerger2007}, self-thermophoresis in 
compressible single-component fluids has also been discussed 
\cite{Brenner2010}. Since in our case, the 
temperature profile depends on the shear rate, we expect the Soret coefficient 
$S_{T}(\dot{\gamma})$ to be a function of $\dot{\gamma}$. 

\begin{figure}[ht!]
\includegraphics[width=.45\textwidth]{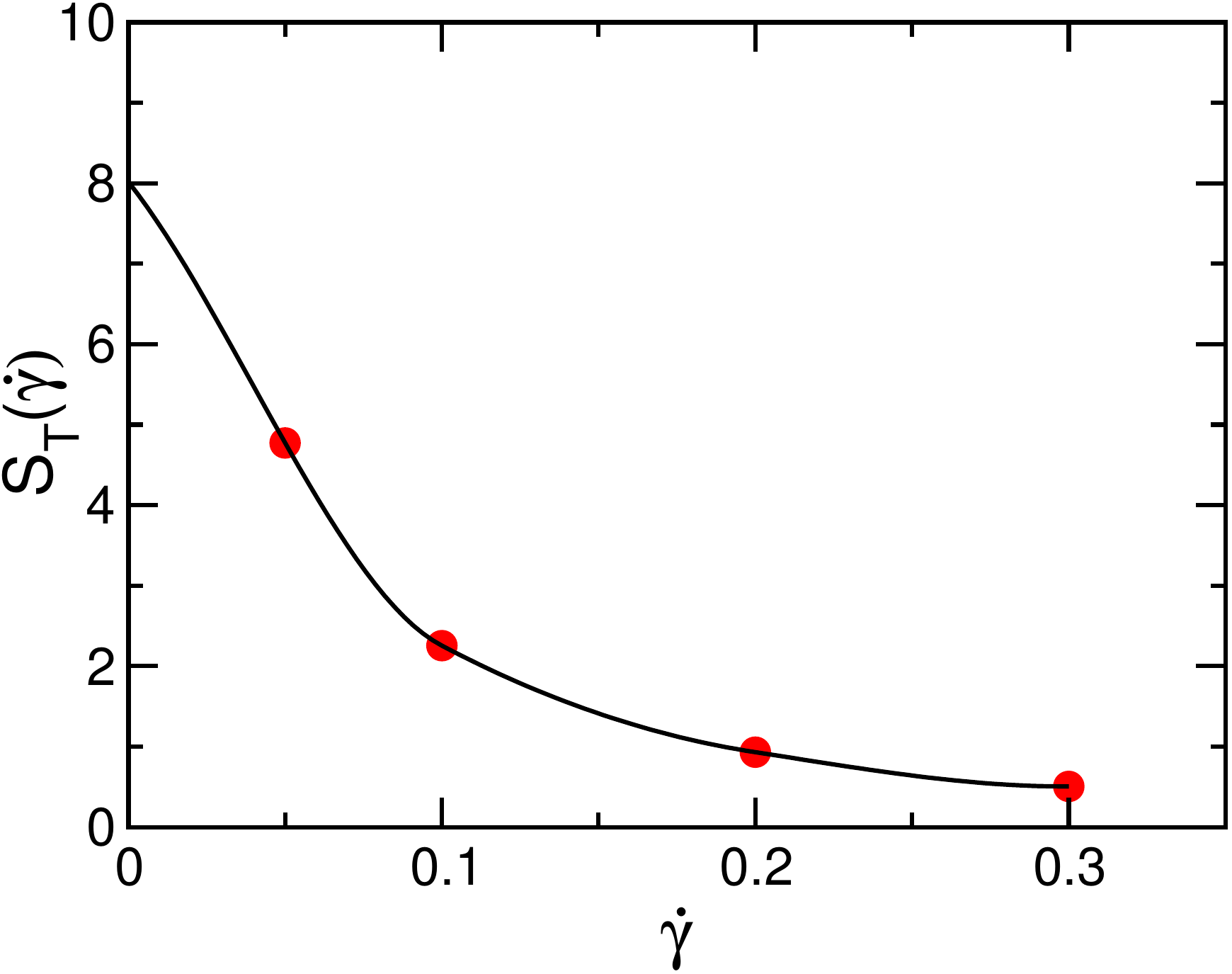}
\caption{\label{soret} The Soret coefficient like parameter is shown for 
different shear rates (solid circles). The solid line is a guide to the eye.}
\end{figure}

In order to calculate $S_{T}(\dot{\gamma})$ , we make use of the 
fact that our compressible fluid can be considered as a binary 
mixture of particles and vacancies. Defining $\rho_b$ as the density profile
at the center of the flow channel $y/L=0.5$ (note that $\rho_b$ is a
function of $\dot\gamma$), the equation that connects the
steady-state distribution of the average density of particles $\langle\rho(y)\rangle$ 
to the steady-state temperature gradient $\langle T(y)\rangle$
is \cite {Gans2003},
\begin{equation}
\label{soretEqn}
\frac{\partial\langle\rho\rangle}{\partial y} = -S_T \rho_b (1-\rho_b)\frac{\partial \langle T\rangle}{\partial 
y}. 
\end{equation}
\begin{figure}[ht!]
\includegraphics[width=.45\textwidth]{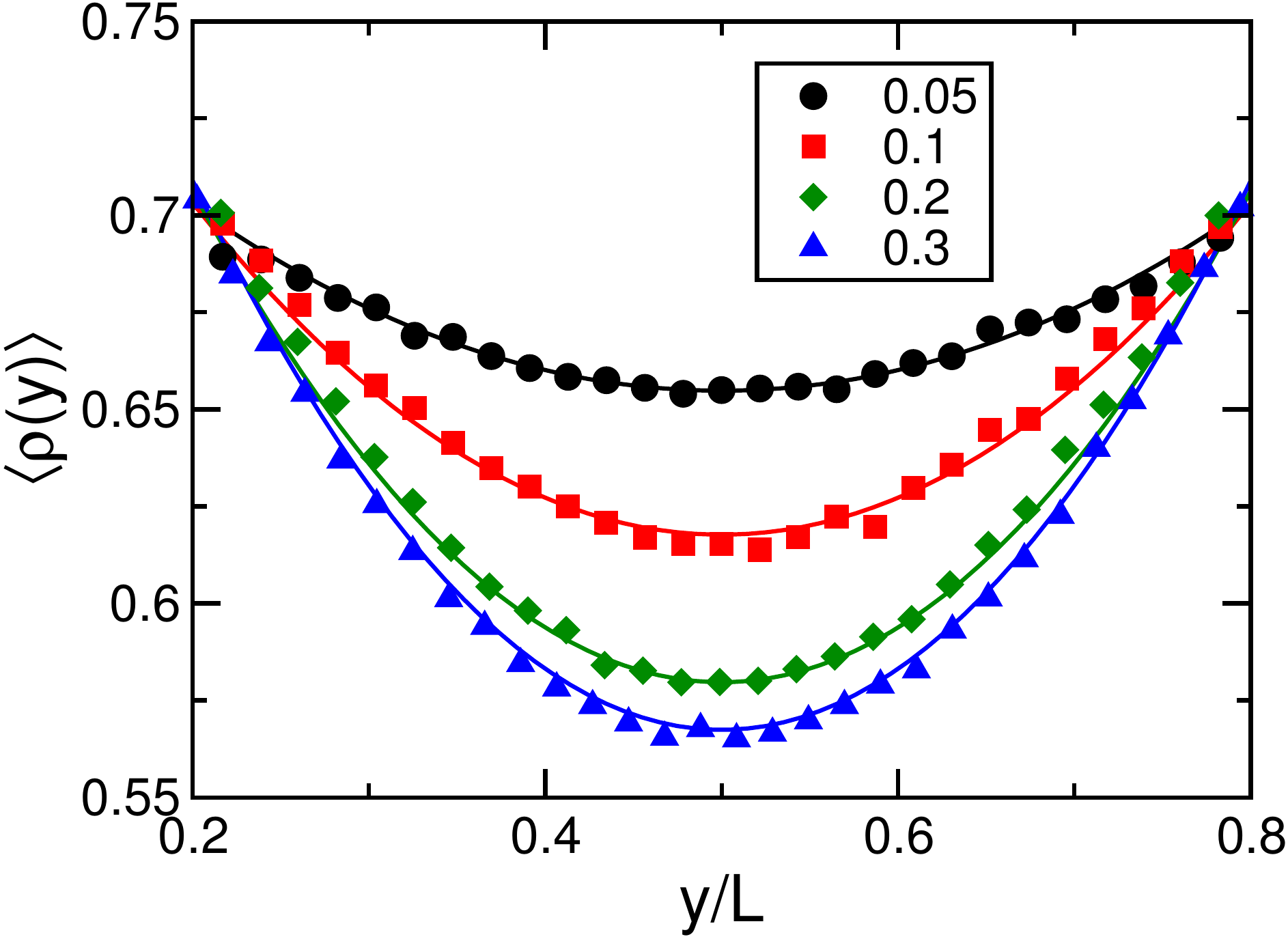}
\caption{\label{densityFit} 
The filled symbols give the density 
profiles in 
the bulk fluid for different shear rates (see legend). The solid lines are 
parabolic fits.}
\end{figure}
Note that while the temperature profile $\langle T(y)\rangle$ is always
parabolic, the density profile $\langle\rho(y)\rangle$ is not 
(compare Figs.~\ref{tempProfile} and~\ref{densityProfileComp}). Upon some reflection
we conclude that the linear response relation Eq. \ref {soretEqn} is valid only in the central
region of the channel, where the deviations from $\rho_b$ are 
small. In Fig.~\ref{densityFit} we show that the measured density profiles 
in the region $0.2\le y/L\le0.8$ away from
the walls,  are indeed parabolic and therefore the density  
and the temperature profiles can be fitted by the quadratic expressions 
$\langle\rho(y,\dot{\gamma})\rangle = 
a(\dot{\gamma})\times(y-0.5)^2 + \rho_{b}(\dot{\gamma})$ and  
$\langle T(y,\dot{\gamma})\rangle = b(\dot{\gamma})\times(y-0.5)^2 + T_{b}(\dot{\gamma})$ 
respectively. In these expressions, $a$ and $b$ are the constants obtained by 
fitting the density and temperature plots with a parabola and 
$T_b$ is the temperature at $y/L = 0.5$. Substituting the above expressions
into Eq. \ref {soretEqn} we obtain the following expression for the 
Soret coefficient $S_{T}$:
\begin{equation}
 S_T(\dot{\gamma}) = -\frac{a}{b \rho_{b}(\dot{\gamma}) [1-\rho_{b}(\dot{\gamma})]}.
\end{equation}
$S_{T}$ as a function of shear rate $\dot{\gamma}$ is 
plotted in Fig.~\ref{soret}.
The equilibrium Soret coefficient ($S_{T} \approx 8$) can be obtained
upon extrapolating the available data points to $\dot{\gamma} = 0$.

Since at high shear rates  large deviations from the linear velocity 
profile are observed, we expect other signatures of non-Newtonian fluid behavior to appear as well
(e.g., a non-uniform distribution of stresses, shear thinning/thickening of viscosity, etc).
In order to compute the stress tensor from our MD results, we express the $xy$ component of the microscopic stress tensor
in terms of  instantaneous particle velocities and interparticle forces:
\begin{equation}
 \label{stressTensor}\sigma_{xy} = \sum_{i=1}^{N} m v_x^i v_y^i + 
\sum_{j>i}^{N} r_{ijx}F_{ijy}.
\end{equation}
\begin{figure}[ht!]
\includegraphics[width=.45\textwidth]{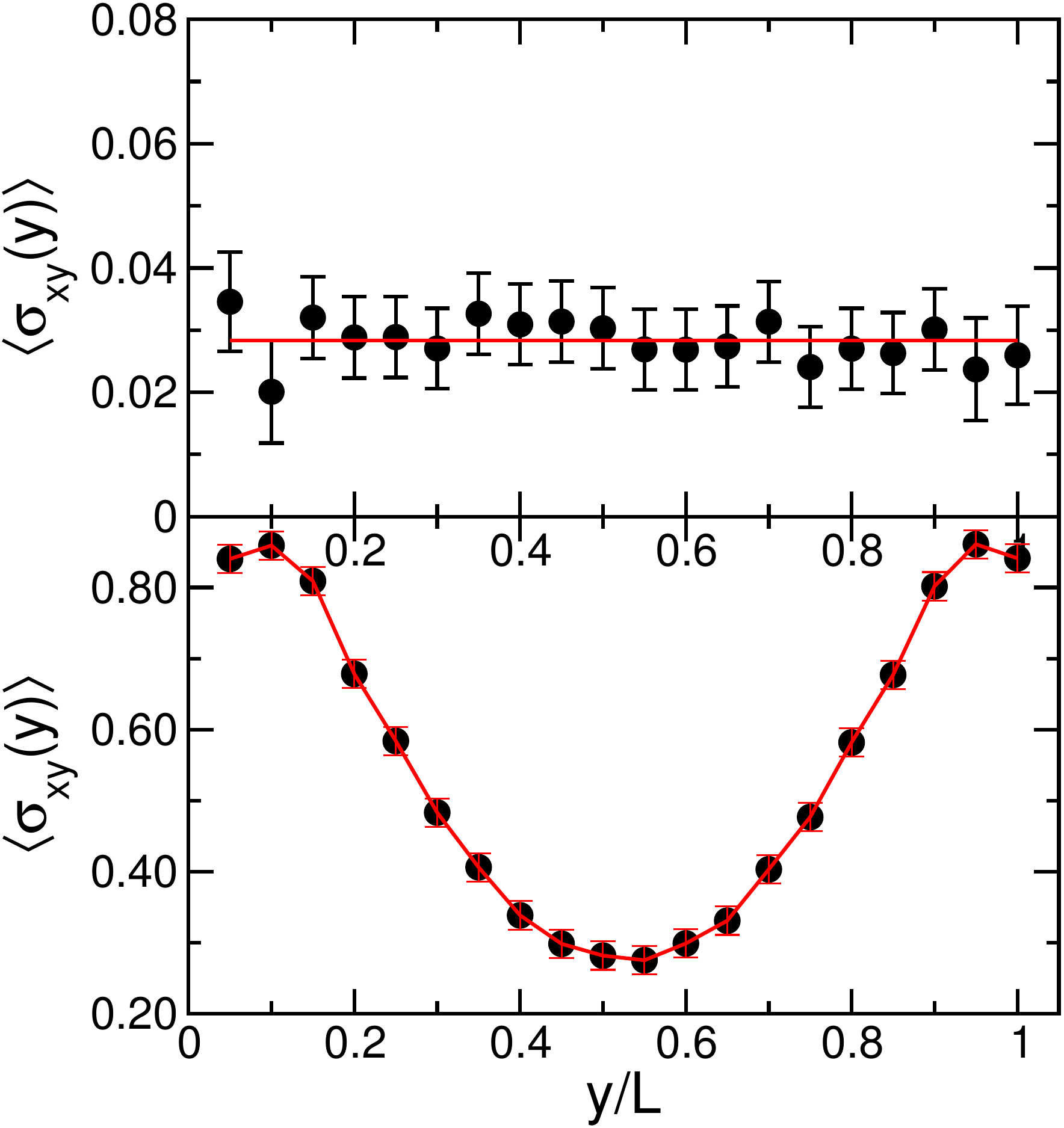}
\caption{\label{sxybin}
 In the upper panel shear stress is shown across $20$ bins in the 
$y$-direction at $\dot{\gamma} = 0.01$. In the lower panel, we show the same 
plot at $\dot{\gamma} = 0.3.$ }
\end{figure}
We then divide the system into $20$ layers (bins) along the 
$y$-axis and average the stress in each layer over $x$ and over
time. As expected, at lower shear rates (e.g., 
$\dot{\gamma} = 0.01$), the shear stress is distributed uniformly perpendicular 
to the direction of flow. At higher shear rates $\langle \sigma_{xy}\rangle$ becomes a function of $y$, with a minimum 
at the center of 
the channel (Fig.~\ref{sxybin}).
\begin{figure}[ht!]
\includegraphics[width=.45\textwidth]{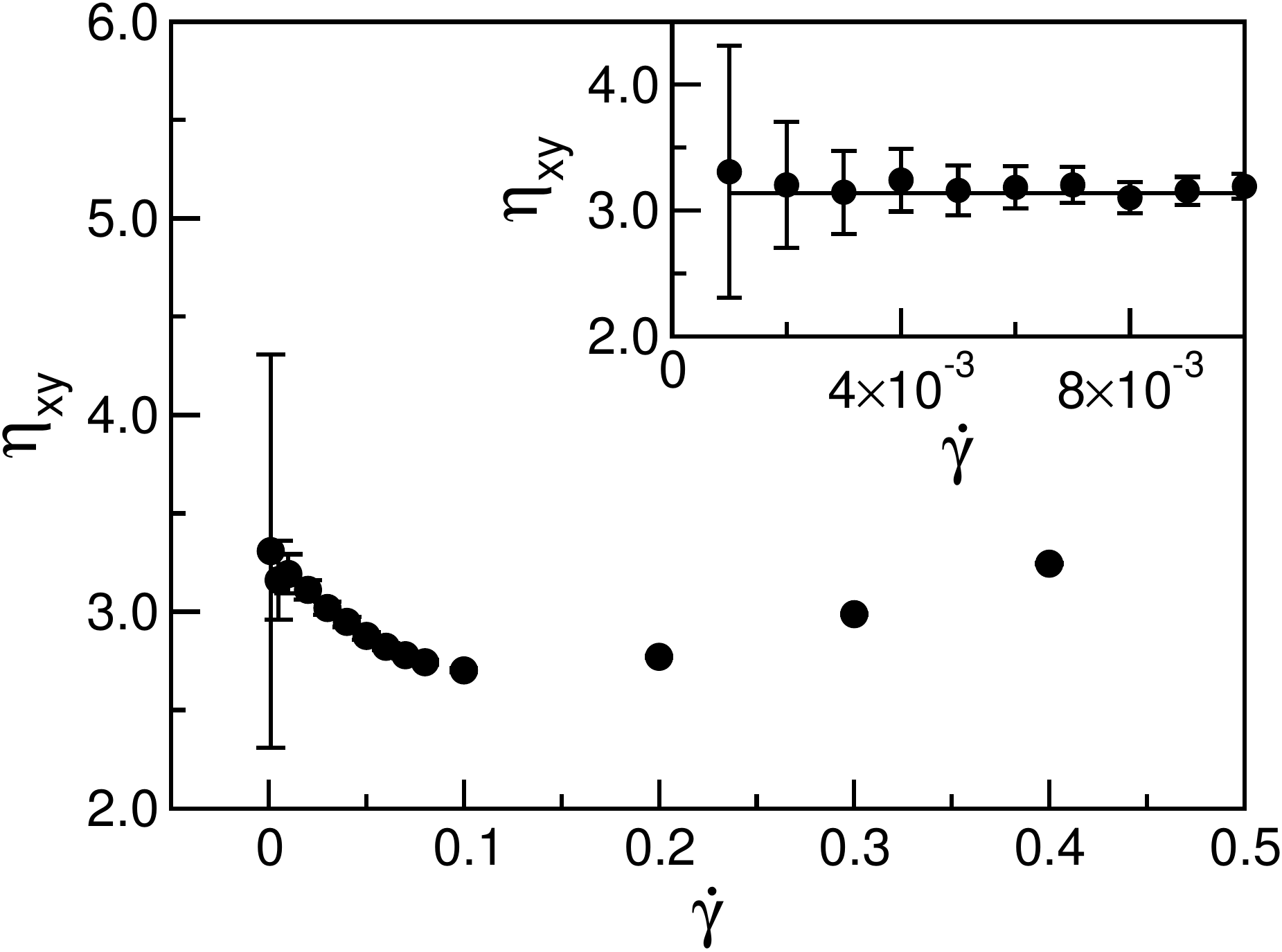}
\caption{\label{shearViscosity}
In the main figure shear viscosity $\eta_{xy}$ is shown for shear rates 
between $0.001$ to $0.4$. In the inset $\eta_{xy}$ for shear rates 
$0.001$-$0.01$ is presented. The range $0.001$-$0.01$ is repeated in 
the inset for clarity.}
\end{figure}

The shear viscosity $\eta_{xy}$ is given by
\begin{equation}
 \eta_{xy} = \frac{\langle\langle\sigma_{xy}\rangle\rangle}{\dot{\gamma}},
\end{equation}
where $\langle\langle \rangle\rangle$ denotes averaging over the volume of the system
and over time. 
In Fig.~\ref{shearViscosity}, we present the shear viscosity $\eta_{xy}$ 
as a function of shear rate $\dot{\gamma}$. Within the accuracy of our simulation, 
the viscosity remains constant 
up to $\dot{\gamma}\approx 0.01$ (see inset in Fig. \ref{shearViscosity}). As the
shear rate is further increased, there is a gradual transition to a shear 
thinning  regime 
in which viscosity decreases with increasing shear 
rate. This  regime extends up to 
$\dot{\gamma} \approx 0.1$ at which point shear thinning is replaced by shear thickening and 
viscosity increases with shear rate.   

\section{Discussion}\label{discussion}
In this paper we used computer simulations to study the dynamics of a 
compressible fluid in steady shear 
flow. We found that as the shear rate is increased the fluid develops 
a highly non-uniform density profile, with pronounced solid-like layering near the walls,
the extent of which increases progressively with shear rate. This shear-induced 
layering  originates in viscous heating of the fluid by the imposed shear
that leads to the appearance of large temperature gradients between the bulk
of the fluid and the confining walls which are kept at constant temperature (for 
technical reasons we thermostat the fluid layers near the walls rather than the 
walls themselves).
The coupling between temperature and density gradients gives rise to the 
Ludwig-Soret effect and has been the subject of numerous
studies in the past but, to the best of our knowledge, the present work is the 
first to demonstrate that such thermophoretic effects can take place in a {\it homogeneous fluid in
shear flow}. 

In addition to the study of the density and the temperature profiles we also 
looked at other dynamical properties of the fluid such as its velocity profile, 
shear stress and viscosity. We observed 
large deviations from linear velocity profiles, and other non-Newtonian phenomena at
high shear rates, such as inhomogeneous stress distributions, shear thinning and shear 
thickening. While locking has been found in previous computer simulations in the limit of large
wall roughness, the observation of shear-enhanced locking has not been reported prior to this work.

All the results reported so far were obtained for a particular value of temperature ($0.44$) and 
density ($0.694$), not 
far from the triple point of the two dimensional LJ system. We studied the behavior of the system
at other temperatures and densities as well (not shown). As expected, we find that if the temperature is raised 
from $0.44$ to $1.0$ at triple point density (0.694) all the effects reported in this work (e.g., layering and deviations from linear 
velocity profile) are strongly suppressed. If the temperature is reduced towards the fluid-solid transition temperature 
of $0.4$, the above effects are strongly enhanced but since under these conditions the lengthscale of density fluctuations 
becomes comparable to system size, we did not undertake a careful study of this regime. We found that both layering and
locking effects can be enhanced  by keeping the temperature constant (at $0.44$) and increasing the density to 0.77
which is close to liquid-solid coexistence density at this temperature. Since the results are qualitatively similar to 
the ones reported in this work, we will not present them here.

We would like to conclude with a comment on the experimental relevance of our results. Even though we simulated the flow of compressible fluids in two dimensions, we expect similar behavior to be observed in compressible near-critical fluids in three dimensions as well. While the shear rates reached in the simulations are unrealistically high, we believe that shear-induced heating of the kind described in our work can be experimentally realized under less extreme conditions in real fluids. Another interesting and experimentally-relevant possibility is that shear-induced thermophoresis can lead to spatial segregation in multicomponent fluid mixtures. MD simulations of such systems are currently under way.

\vspace*{0.2cm}

\begin{acknowledgments}
The authors gratefully acknowledge fruitful discussions with D. Osmanovic and 
D. C. Rapaport. YR's work was supported by a grant from the Israel Science Foundation.
\end{acknowledgments}

\end{document}